FOUR LAST "CONJECTURES"
Philip W Anderson, Princeton University (emeritus)

Abstract: I collect here some ideas on four topics which have not gained much acceptance in the community; hence, "conjectures."  The first three of these are in my area of expertise, and   I personally consider them to be verified, if by evidence some others might find less than convincing;  on the the fourth I do not claim expert knowledge but it has not been refuted in any way that I understand. The first is the proposition that the ground state of a solid made up of simple bosons is phase coherent and sustains "overlap currents";  the second proposes that  the cuprate superconductors' upper phase transition is into an Ong vortex liquid phase which quantizes flux and  consequently exhibits a characteristic singular, nonlinear diamagnetic behavior;  the third proposes that the metallic phase of the doped Mott insulator, if it has a Fermi surface, has deep entanglement and an orthogonality catastrophe causing a Fermi surface singularity. This manifests itself as the "strange metal" of the cuprate phase diagram. The fourth conjecture is that the "Dark energy" in cosmology models is the consequence at least in part of gravitational radiation carrying energy past us.

INTRODUCTION
As I reach my later 90's I find that I am increasingly preoccupied with unfinished business from my final decade as an active physicist.  One of my earlier contributions, which I had considered to be rather solidly reasoned, was described by a famous mathematically inclined colleague as a "conjecture", and as such I will label the four leftovers from my scientific career which I here describe.  As far as I am concerned, each (except perhaps for the fourth, where I am venturing outside of my area of expertise) is reasonably soundly based in theory, and has some experimental evidence in its favor; but the various communities into which they have been introduced have, almost to a man, ignored or in some cases rejected them. I think in each case they have some significance for the fundamentals of the phenomenology and they belong in the textbooks, which is why I keep worrying about them.

I. Overlap currents and NCRI in the Bose solid ground state.
II. The Diamagnetic response of the vortex liquid (and its existence in the cuprates)
III. The "hidden" Fermi surface in projective Fermi liquids—connection with chiral anomalies and anomalous exponents.
IV. Gravitational radiation as dark energy.

I. Overlap Currents

I learned, practically in my cradle (actually from Bill MacMillan's thesis), that the ground state wave function of a system of mutually repelling bosons should necessarily be real and positive, a fact which made his early Monte Carlo simulations infinitely easier. This is, presumably, equally true of the quantum solid as of the liquid.

But the quantum solid is presumed to have (and actually does have, in the case of solid He) "Mott insulator" status, in the sense that its atom sites are each commensurately occupied by a single boson, so that the phase of the wave function is apparently meaningless: $\delta n=0$ seems to imply $\delta\phi=\infty$, as is trumpeted by the theoretical group at UMass, and echoed by various other theorists. So the phase should not matter—the calculations may be just more convenient ignoring phase.

But is δn=0?   The simplest approximation to a possible ground state wave function is a Hartree product of  localized Boson wave functions,

$B^*_i = \int \psi^*(r) f(r-R_i) dr$;   $\Psi = \prod_i B_i^*$

where f is some appropriately localized function around each site of a lattice $R_i$, presumably calculated selfconsistently a la Hartree.  In the case of Fermions, the Hartree-Fock procedure automatically produces orthogonal localized functions $f_i$, and the many-body wave function is a Slater determinant of these.  Therefore, both the individual functions, and the many-body determinant, must have sign changes if they overlap even slightly.

But, as I pointed out in my book, there is no corresponding requirement that the f's for *bosons* should be orthogonal, and in fact they cannot be and maintain the sign rule for the ground state, unless they are unrealistically confined to separate, disparate regions around each site.  If they are not orthogonal, there is interference between the bosons and there are accompanying number fluctuations.

Actually, most of the numerical calculations on helium do not use the Hartree  method sketched above; rather they model the wave function as a product (actually a permanent) of Bijl-Jastrow factors for each pair of particles, choosing each function variationally. As MacMillan pointed out, this leads to the exact partition function for classical particles,  since the Hamiltonian is a sum of pair interactions.

Although these methods lead  quite successfully to numerical results  in agreement with experiment, they leave us a bit bewildered in deciding  how to deal with the phase of the wave function—it is necessarily assumed to be zero. If the energy depends on the phase, as Kohn has pointed out, the system is *not* an insulator.

To deal with the phase, we need to go back to the Hartree-like product representation. This is a natural representation to use for the Bose-Hubbard model, which ignores the phonon spectrum, but not for the true solid. On the other hand, I showed in my messy chapter 4 that the phonon spectrum is perfectly compatible with a representation based upon local wells—the phonons may be made up out of linear combinations of exciton-like excitations of the local functions. Such excitations carry no current and are irrelevant to the phase question, since they do not carry a site-dependent phase.

Since the phonons are irrelevant to the question we are asking, the Bose-Hubbard model becomes a realistic starting point for continuation via a perturbation theory in t/U, which starts from a classical –like solid and allows the kinetic energy t to grow—ie basically an expansion in h. Monien and Elstner[1] have developed such a perturbation theory for the 2D B-H model, which, they demonstrate, converges, and in the lowest nontrivial order its solution turns out to be the product of nonorthogonal bosons; in higher order its structure is more complex but there is no indication that this complexity could restore orthogonality or reduce the interference between the local bosons.

It seems irrefutable that there is, first, a meaningful local phase; and second, that there are "superexchange" interactions between the phases on the different sites. Ginsburg and Landau confronted the problem of what form such an interaction might take for liquid He 4, but for our discrete case we use $-J_{ij}\cos(\varphi_i-\varphi_j)$, a Heisenberg –type model. Such an effective Hamiltonian leads to long-range phase condensation in 3D, topological order in 2D, at a temperature of the order of J. It leads to currents $J=\nabla\varphi$. The restriction to Mott insulator status is enforced by requiring $\nabla\cdot J=0$.

The excitation spectrum certainly has a gap for particle and hole excitations—it's a Mott insulator. It also has a gap, the "core energy", for zeroes of the phase field, but that will be much smaller. These singularities of the phase field will serve as the cores of vortex excitations, which in 2D will proliferate above the phase transition, destroying the long-range topological (in 2D) order .

I would like very much to identify the transition to the phase-ordered state with the "giant plasticity" transition observed below 10 mdeg K in a very pure crystal of solid He by Haziot et al.[2] It was a sudden increase in a shear modulus of about 50% at or below 10 mdeg K. Significantly, the increased shear modulus is that predicted by the Bijl-Jastrow computations, which predicate phase coherence because they assume a real wave-function.

Nowadays it is also experimentally possible to investigate the 2D Bose Hubbard model using cold atoms, and as I have mentioned the experimental phase diagram is in agreement with perturbation theory as to the extent of the Mott phase, but experiments have not studied its dynamic response, and are not yet, probably, at low enough temperature to exhibit the phase coherence. I think the experiment of investigating the dynamic response at low temperatures is urgently needed; and it should not be too difficult if carried out just below the Mott transition, where NCRI should be quite appreciable.

i think this conjecture is of some philosophical interest because it probably applies to most crystalline solids, if at unrealistically low temperatures: the solids we work with are not in their ground states.

II. The Vortex Liquid response

The thesis of Yayu Wang (Princeton, 2004), written under N P Ong, contains a library of instances of the Nernst effect exhibited near the superconducting-"normal" phase transition of several cuprate superconductors.  From the start Ong and his students attributed the large Nernst effects seen to vortex motions, since they appeared essentially continuous with those seen below this phase transition where the material is known to be in a vortex (flux line solid) state; and the heat transport in the Nernst effect is known to be carried by the motion of quantized flux lines.  A second student , Lu Li, and Ong extended these measurements , and also added measurements of the diamagnetic  response over an even larger range of cuprate specimens.  Diamagnetism is a direct, thermodynamic measurement of the energy in the structure of quantized flux lines, and it was found that the transport measurement of this energy via the Nernst effect and the direct measurement via the diamagnetic susceptibility agreed in every case where the comparison was made.

Quite early in the history of attempts to understand the cuprates (1995) Emery and Kivelson [3] proposed that the superconducting phase transition, on the left side of the "dome",  is a Berezinski-Kosterlitz-Thouless vortex proliferation in the two-dimensional planes of the cuprates; this indeed gave a quite good quantitative fit to the transition temperature, and placed the putative BCS pair-breaking temperature quite a bit higher than the observed Tc , in agreement with estimates by Rice et al of 1988 [4]and later. In fact, already at that time we had a reasonable quantitative theory of the transition temperature, if we had only trusted it.

When I examined the energy of the vortex structure calculated by Kosterlitz and Thouless

$$E/2\pi = \sum_i q_i^2 \ln(R/a) - \sum_{i \neq j} q_i q_j \ln R/r_{ij}$$

$$= (\sum_i q_i)^2 \ln(R/a) + \sum_{i \neq j} q_i q_j \ln(r_{ij}/a) \qquad (4)$$

(multiplied by a constant factor $(h^2/m)\rho_s$)
and used in their theory of the loss of phase coherence, I found that the transition did not totally disorder the phase. The superfluid density remains finite throughout the vortex proliferation and the self-energy embodied in the first term in the second line cannot be screened out, counterintuitively: the energy increase resulting from adding an unpaired vortex cannot be affected by any rearrangement of the rest of the vortex structure. In that it doesn't depend on the coordinates $r_i$ of the vortices. This term, however, is crucial to the Nernst effect and the diamagnetism.

It is valid to say that the BKT transition is between two topologically ordered phases, one –the lower—which behaves like a true superconductor because the phase is locked, and the vortex structure if present becomes a three-dimensional regular array. The second, higher, one has still got a superfluid density, and therefore an energy which is logarithmically divergent in the sample size is required to add the first quantum of vorticity, but otherwise the phase fluctuates.

If we now apply a magnetic field containing more than one flux quantum, the energy does not continue to rise quadratically with the number of unpaired vortices but distributes these uniformly so as to mimic the uniform vorticity of a conventional diamagnet. At scales larger than the average intervortex distance $r_B=\sqrt{(\Phi_0/B)}=2e/B\sqrt{(2\pi hc)^{-1}}$, the currents due to the diamagnetic response to the magnetic field effectively cancel the vorticity currents and no extra energy is required; but below this scale, the vorticity current is granular and requires a self-energy for each vortex. This self-energy is just the self-energy term with the upper length cutoff set at $r_B$. and this, then, is the energy associated wiith the diamagnetic currents, which is measured by the vortex Nernst effect or by the diamagnetism itself.

$$E = consts \times B \int_{\max[T, \sqrt{B/K}]}^{\Delta_{max}} p_0(\Delta) d\Delta \ln(K\Delta^2 / B)$$

It is this form which we see repeatedly in the experimental data on essentially all of the cuprate superconductors. Here K is a constant which is defined by $K\Delta^2 = H_{c2}$ and $p_0$ is the distribution of gaps in energy. The lower integration limit is set by the fact that either thermal fluctuations or rapid precession can make the gap ineffective: it disappears from the low side as temperature or field is increased. The result of a sample calculation is shown in a figure, and a typical experimental result in another, from Lu Li's thesis.

Princeton, (2005)

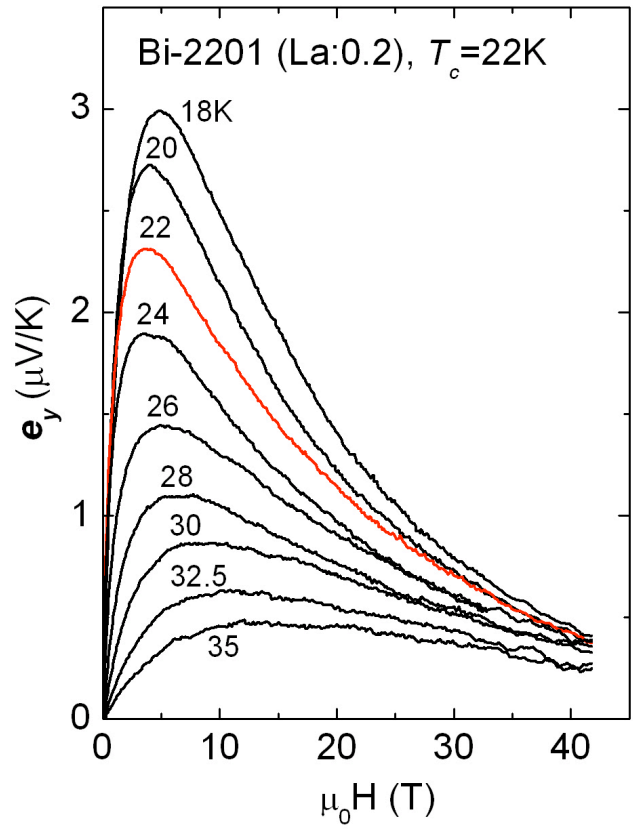

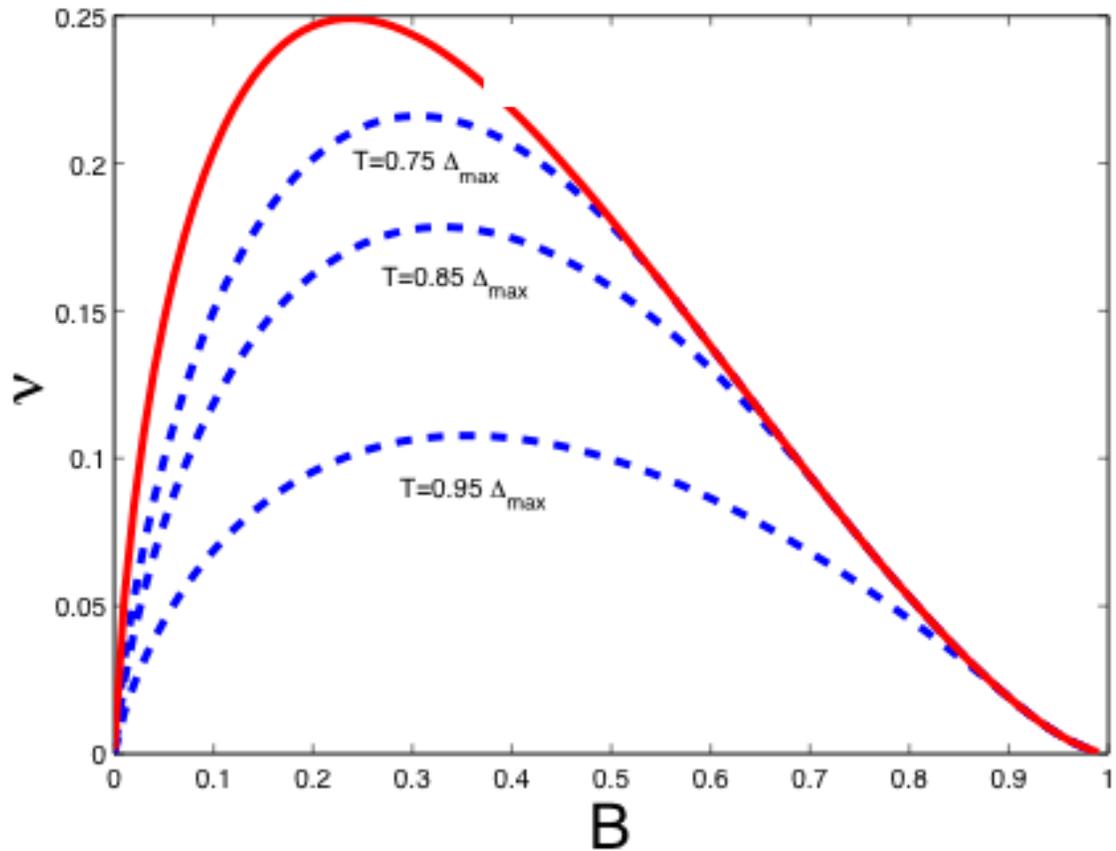

A number of people have complained that I am drawing conclusions simply from a general "resemblance" of the nonlinear magnetic response to the theoretical curve rather than from precise curve-fitting technology. My defense is that no one else has made any effort to understand or to explain this very characteristic response in any realistic way, and that it is of a universal and rather unexpected form. In particular, that the susceptibility approaches the origin in field with a clearly logarithmic behavior, though obvious experimental difficulties prevent actually exhibiting the singularity, is very hard to understand except as a vortex effect. It is almost equally strange that the other singularity, at Hc2, is devoid of fluctuation effects and is simply linear. Of course, the associated evidence that the superconducting phase transition is BKT and that the Ong phase does exist and is nearly universal is very strong—for instance, the recent measurement s by Sebastian et al demonstrating that at very low temperatures where all vortex flow is pinned zero resistance persists up to Hc2.

III The hidden Fermi Liquid: Anomalies due to Projection

At the "Woodstock" special session of the "March" meeting of the American Physical Society, over 30 years ago in 1987, which signalized the near-hysteria over the discovery of 90 K superconductivity, a flyer was passed out for the attendees to sign and commemorate their attendance.  On the flyer was a simplified sketch of the resistivity vs temperature curves that the tens of experimentalists would have reported on their miscellaneous samples of YBCO (yttrium barium copper oxide) and other cuprates.  The resistivity remained zero up to the then startlingly high Tc; then it leapt up to a finite, not low, value *and continued not on a level, but rose linearly with the absolute temperature in degrees K!*
This was not merely a casual sketch—it is a reasonably accurate description of the actual data, and it immediately grabbed my interest, since it suggested that the metal was not formally a Fermi liquid—since all of the lore of the Fermi liquid relies on the idea that because of the exclusion principle, the scattering of the quasiparticles becomes negligible compared to their energy as they approach the Fermi surface. Linear in T means that the scattering rate is of exactly the **same order as the energy**.  This is the marginal case, and indeed one of the (unsuccessful) theoretical proposals which appeared was called the "Marginal Fermi Liquid Theory".  I cannot criticize such efforts too severely because I followed my own ineffectual efforts to  explain this vainly for more than a decade. To me, and I suspect many other theorists, the real excitement in that sketch lay not in the high value of Tc, which was easily and immediately explained as the result of large exchange interactions(though not so easily accepted by the diverse community of specialists which grew up)—but the many hints that a new paradigm for metals was necessary.

This linear dependence of the resistivity (and hence of the relaxation rate $1/\tau$) on temperature was not the only odd feature of the simple transport properties. Most striking was the response to a magnetic field—the Hall effect. The most general way of describing this is just that there is **simply no evidence of the linear T behavior in the relaxation rate of the Hall current**. In typical cases, the Hall effect became T-dependent, varying roughly as $1/T$, and the relaxation rate is a conventional $T^2$.

For 30 years, these observations have held up and been repeated more times than I can recall. **There is every evidence that as far as currents parallel to the Fermi surface are concerned, there is a conventional Fermi surface, and there is no linear T term..** More recent experiments have even found that magnetooscillations can be seen in some cuprates in high fields.

A second embarrassing fact is that the linear rise in resistivity continues on right past the "Joffe-Regel limit" set by conventional transport theory, without a break.

These anomalies have not been addressed by **any** of the many alternative theories—for instance, the many papers ascribing "linear T' behavior to the penumbra of some mysterious "quantum critical point". All of these anomalies are directly and simply explained by the HFL idea.

My first correct insight came from examining the data on the infrared conductivity, first, by Schlesinger and Collins, and then of Nicole Bontemps' group[5], and realizing that they were best represented by a power law; the "marginal" theory had attempted to fit them with a logarithmic behavior, but in fact better analysis showed one that the "Linear T" implied a complex conductivity with a constant "loss angle", which behavior describes a power law.

A nearly universal fixture of the "Mott physics" of the strongly interacting Fermi systems which we were using to describe the doped cuprates is the "upper Hubbard Band", which is often seen optically, (but not universally because it can lie above other conduction bands), and invariably appears in computer studies of the strongly interacting Hubbard model which, we had come to feel, contained the essential physics of the cuprates. This band is visualized as being constructed of states which my student Ted Hsu called "antibound": doubly occupied states in the topmost of the dp hybrid bands, and it is gapped above that band if the Mott interaction U is above a critical value, which I think is quite generally exceeded for the cuprates. ( My opinion is not universally shared, but that's irrelevant.) At that critical value, the particle-hole ladder diagrams of perturbation theory diverge so in fact one may argue that there **is** a critical point in the problem; the singular behavior, however, **does not arise from fluctuations about that critical point but from the nature of the fixed point** to which it leads. (This feature it shares with conjecture II). And that fixed point must describe a band which does not contain the antibound states which have split out from it and formed the upper Hubbard band. In working with the conventional band, we are not "playing cards with a full deck": the band must form itself in a restricted Hilbert space The nature of the fixed point is much more easily established by situating oneself near it than it is by futile attempts to gently raise the interaction and sneak past the barrier between phases. This is one way of seeing the nature of the "hidden Fermi liquid" theory which I produced: we assume that the U interaction is quite large and that the antibound states which we must project out of the Hilbert space are quite close to simple doubly-occupied states, just as the Fermi liquid theory models the interacting theory on a perfectly free one. We are by no means assured that the quantitative results will be perfect but the nature of the beast will be the same. The upper Hubbard band is presumed to be made up of all of the doubly-occupied states, and we project all these out of our Hilbert space—the equivalent of driving a Hubbard U term to infinity.

The result is the t-J Hamiltonian; J is the residual "superexchange", which would vanish if U were actually infinite: it is the second-order consequence of the kinetic energy.

$$H_{t-J} = P \sum_{i<j,\sigma} t_{ij} c_{i,\sigma} * c_{j,\sigma} P + \sum_{i,j} J_{ij} S_i \cdot S_j$$

P is the Gutzwiller projector ; it is not necessary to apply it to the J term, through which it commutes, since exchange does not alter occupancies.

$$P = \prod_i (1 - n_{i,\downarrow} n_{i,\uparrow})$$

At this point, if I were to continue with the exposition of the theory I would be self-plagiarizing, so I refer the reader to my article Phys Rev B 78, 178505(2009) . In that article I observe that there are two kinds of Fermion excitations of the projected Hamiltonian, which I call "quasiparticles" and "pseudoparticles." There are Fermion excitations which are exact eigenexcitations of the Hamiltonian but do **not have** any overlap with a single introduced particle (or hole), because of the "orthogonality catastrophe", as I named it long ago: they have Z=0. These excitations have a Fermi surface at which the scattering vanishes because of the exclusion principle. It increases with energy quadratically; this Fermi surface has the Luttinger volume because it is simply the continuaton from below the critical U, and the antibound states of the uhb are not near the Fermi surface and do not affect scattering singularly. These pseudopartcles precess in a magnetic field precisely as conventional quasiparticles do, since the projector commutes with magnetic field, so that the hall angle (which will be determined by the pseudoparticles) acts perfectly normally as is observed in many surveys, for instance that of Segawa and Ando, Phys Rev B 69,104521 (2004).

I define "quasiparticles' as the result of acting on the physical system with a particle or hole creation operator $c_{i,\sigma}^*$ or $c_{i,\sigma}$. Thus where in creating the pseudoparticle one is creating an excitation wholly **inside** the projected subspace, in creating the quasiparticle one realizes that adding an extra particle modifies the projector by modifying the space in which it acts. I pointed out that it was possible to write the kinetic energy entirely in terms of a modified set of Fermion operators, which I decorated with "hats": to quote myself,

We designate the "real" Fermions which represent physical creation and destruction operators acting in the projected subspace by "hat" operators which do not create or destroy any doubly-occupied sites. These are easily seen to be

$$\hat{c}_{i\sigma} = (1 - n_{i-\sigma})c_{i\sigma} \quad (\hat{c}^*_{i\sigma} = c^*_{i\sigma}(1 - n_{i-\sigma})) \quad [11]$$

Here the hatless operators are to be thought of as operating within the unprojected space, that is they operate on the hidden Fermi liquid.

The Green's function for inserting the hole at time 0 and removing it at time t, then, might be written as

$$G_{ii}(0,t) = \langle 0|\hat{c}^*_{i,\sigma}(0)\hat{c}_{i,\sigma}(t)|0\rangle$$
$$\cong \langle 0|c^*_{i,\sigma}(t)c_{i,\sigma}(0)|0\rangle G_{-\sigma}^*(t)$$
$$= G_0(t)G^*(t) \quad \text{where} \quad [12]$$
$$G^* = \langle 0|(1-n_{i,-\sigma}[t])(1-n_{i,-\sigma}[0])|0\rangle$$

here $G_0$ is the free-Fermion Green's function, which for the single-site case of [12] is proportional simply to $1/t$.

I was a bit remiss in not discussing the justification for factorizing the three-fermion operators in the way that I do. This is essentially placing the model system as near to the fixed point as possible, where the higher–order effects of t are minimized. There are actually two channels which lead to a single particle and a projector, the obvious one and one in which the original projector supplies an opposite-spin fermion, leaving behind a projector which removes one of the three components of the spin. In the absence of J, we may choose any axis we like and the two channels are equivalent; but in the real cuprates, J is quite large and the spin channel may—and does—exhibit additional features. These are the structures in the tunneling conductivity which some authors have attempted to identify as Rowell-McMillan bumps and .interpret accordingly. But the essential nature of the insertion of a quasiparticle is that it is accompanied by a quantum quench acting on the opposite-spin Fermi sea, and because of the profound entanglement of the Fermi sea system the response to the quench is singular and not meromorphic, as was first foreshadowed in my work in the late 1960's and subsequently on what I called the "Infrared catastrophe". The response G* is the overlap between the projected state and the state before projection, which must be zero at zero energy because of the catastrophe: $G^* \propto \omega^p$.

Thus the underlying Fermi surface is not a step-function but a power law singularity. This "catastrophe"(which is closely related mathematically to the field theorists' "anomalies") is the source of the anomalous exponents which characterize the cuprates. It is **not related to fluctuations about any quantum critical point;** it is a fixed-point property of a distinct phase, the Mott or t-J metal, of which the cuprates are the simplest exemplar.

Thus the two distinctive features of the resistivity behavior shown in the 1987 diagram at "Woodstock" are indeed deeply related: the high transition temperature results from the strong antiferromagnetic "super" or "kinetic " exchange interaction which can result from perturbatively removing matrix elements which cause double occupancy, while the anomalous resistivity results from the deep entanglement of the projected subspace. One very important thing to realize is that **in the absence of projection there is no attraction causing singlet pairing**, so that the phase with d-wave pairing is not continuable from any Fermi liquid.

In recent years there has grown up a school of theorists who consider the anomalous behavior of the cuprates as caused by a quantum critical point of some sort. As far as I know no meaningful fit to experiment has ever been achieved this way and I believe this whole literature is without relevance to the physical cuprates.

There is an exception which I cannot condemn as categorically: the scheme of Rice, Robinson and Tsvelik.[6] These authors do not mischaracterize the phenomenon by ignoring the Hall data, and their fit is, like ours, to the sum of two conductivities. One of these is that of the Fermi surface, while the other is that of a one-dimensional Fermi gas which is established by a lattice distortion caused by a divergence of umklapp interaction diagrams.  Such distortions have been seen in some underdoped samples but I have seen no evidence that they persist to the high dopings and energies at which the linear-T mechanism is seen.

IV.Dark energy as gravitational radiation.  The following should be considered as much more amateurish than the three previous conjectures,, which are the result of decades of study of the relevant problem and particularly of the extensive experimental evidence in each case.

In the recent observations of gravitational radiation from a black hole collision event it has been calculated that the mass of the resulting aggregate is several solar masses (about 3, I believe) less than the sum of the masses of the original pair, and that therefore the shell of  gravitational  radiation carries the equivalent energy as it passes us.  If we were now to observe the remnant, we would see it as lighter by 3 solar masses,  and if we were to calculate the local matter density of that sector we would have to conclude that $3p + \rho$, the source term for the gravitational field,  has been lowered by that amount.  The radiation from this event has not added to the gravitating density of somewhere else in the universe, as far as its effect on us is concerned,  because it has passed us at  light speed and is not in the observable sector of the universe for us.  It was not scattered during the large fraction of the Hubble time that it was on its way to us and we must assume that it will not be further scattered or help constitute a gravitational "afterglow" that we can observe. We have to conclude that, however slightly, there is that much less net attraction and that therefore our expansion is slowing down less, i e we are accelerating. This last point is the crux of the argument: the radiation does not conserve total mass from our point of view; the mass is irreversibly lost to the part of the universe which has observed the event.

Now this event, and the total history of events of this particular sort, may not matter very much to the cosmological equations, because these events are relatively rare. But we know that an appreciable fraction of the mass of many galaxies is in the central black hole, which must have been created by similar events which radiated away some appreciable fraction of the original mass. These black holes must be lighter in this sense than the total mass that has gone into them. In fact, the process of generation of gravitational radiation has been going on everywhere in the universe at all times-- for instance, the pulsar pair which was discovered by Taylor and Hulse has been measured to be continually radiating away its gravitational potential energy. An object has recently been tentatively identified as a black hole resulting from the collision of two galaxies and their concomitant massive black holes, and I have seen no estimate of the past frequency of such events. The point is that all of this radiation is created by irreversible processes which dissipate energy which, as far as measurements here and now could ascertain, does not contribute to the net gravitational self-attraction of the universe as we would estimate it. The observable universe is becoming lighter at some unknown rate, depending on how much is being irreversibly radiated away. An appreciable amount of gravitational potential energy would seem to have been radiated away irreversibly in the course of star, galaxy and black hole formation . This does not seem to be accounted for in the present cosmology, and may be a part, or even the whole, of the "dark energy" that is now postulated.

I have not been able to convince myself whether or not any part of the spectrum of electromagnetic radiation has the same irreversible effect. The CBR is presumed to be in thermal equilibrium so these considerations don't apply; but individual events such as supernovae are a different matter.

In ny case, there is a grear deak if gravitational energy missing, and we need to know how much of it is radiated away and constitutes the background which we are just beginning to be able to measure.

,

attempt to fit these and other data with quantum criticality was published by D van der Marel in 2003

[6] **TM Rice**, NJ Robinson, **AM Tsvelik**. Physical Review B 96 (22), 220502(R), 2017.